# Earthquake prediction analysis: The M8 algorithm

G. Molchan and L. Romashkova

*International Institute of Earthquake Prediction Theory and
Mathematical Geophysics, Russian Academy of Science, Moscow, Russia.
The Abdus Salam International Centre for Theoretical Physics,
SAND Group, Trieste, Italy.
E-mail: milchan@mitp.ru, lina@mitp.ru*

**Abstract**

The quality of space-time earthquake prediction is usually characterized by a two-dimensional error diagram $(n,\tau)$, where $n$ is the rate of failures-to-predict and $\tau$ is the normalized measure of space-time alarm. The most interesting space measure for analysis of a prediction strategy is the rate of target events $\lambda(dg)$ in a sub-area $dg$. In this case the quantity $H = 1-(n+\tau)$ determines the prediction capability of the strategy. The uncertainty of $\lambda(dg)$ causes difficulties in estimating $H$ and the statistical significance, $\alpha$, of prediction results. We investigate this problem theoretically and show how the uncertainty of the measure can be taken into account in two situations, viz., the estimation of $\alpha$ and the construction of a confidence zone for $(n,\tau)$-parameters of the random strategies. We use our results to analyse the M8 earthquake prediction algorithm.

## 1. Introduction

To characterize the quality of space-time prediction of $M \geq M_+$ earthquakes (prediction of the yes/no type, see *Molchan, 2003*) two quantities are commonly used, namely, the rate of failures-to-predict $n$ and the rate of space-time alarm:

$$\tau_\omega = \int_G \tau(g)\omega(dg). \tag{1}$$

Here $G$ is the area of prediction, $\tau(g)$ is the rate of alarm at the point $g$, and $\omega(dg)$ is the normalized measure on $G$, i.e., $\int_G \omega(dg) = 1$. The choice of the measure depends on the prediction goals in mind and generally requires special motivation (see e.g., *Kossobokov, 2005; Holliday, et al., 2005; Kagan, 2009a; Zechar & Jordan, 2008*). Among all possible $\omega(dg)$ the most important for application is the normalized rate measure of target events

$$\omega(dg) = \lambda(dg \mid M_+)/\lambda(G \mid M_+), \tag{2}$$

where $\lambda(A/M_+)$ is the rate of $M \geq M_+$ events in $A$. We know that it is only in this special case that all trivial or random guess strategies are represented by the diagonal $D: n+\tau_\omega =1$ of the square $0 \leq n$, $\tau_\omega \leq 1$ (*Molchan & Keilis-Borok, 2008*). In other cases the trivial strategies form a convex neighbourhood of the diagonal where they are arbitrarily mixed with non-random strategies.

In the case (2), the $\tau_\omega$ and
$$H = 1 - (n + \tau_\omega) \tag{3}$$
acquire a simple statistical meaning: they give the rate of randomly and non-randomly (if $H > 0$) predicted target earthquakes, respectively. Therefore, we have $H = 0$ for any trivial and $H = 1$ for an ideal (100% successes) predictions. The intermediate values $0 < H < 1$ determine the degree of non-triviality for a strategy, $\pi$.

The quantity $H$ can also be interpreted as a relative distance of $\pi$ from the set of trivial strategies. To see this, we represent the space $G$ as a set of nonintersecting sub-areas $\{G_i, i = 1,...,k\}$ where an alarm applies at once to all points of the sub-area concerned. Then the prediction characteristic $\{n, \tau(g)\}$ can be treated as the vector $(n, \tau_1,...,\tau_k) = \vec{\varepsilon}$ with $\tau_i = \tau(g)$ for $g \in G_i$, and the trivial strategies as points of the hyperplane $\tilde{D}: n + \sum_{i=1}^{k} \tau_i \omega_i = 1$, where $\omega_i = \omega(G_i)$ and $\omega$ is given by (2). It is easy to check that
$$H = \rho(\vec{\varepsilon}, \tilde{D}) / \rho(\vec{0}, \tilde{D}), \tag{4}$$
where $\rho(\vec{\varepsilon}, \tilde{D})$ is the Euclidian distance between $\vec{\varepsilon}$ and $\tilde{D}$, and $\vec{0} = (0,...0)$ is the parameter vector of the ideal strategy. The representation (4) evidently remains true on the $(n, \tau_\omega)$ plane; in this case $\vec{\varepsilon} = (n, \tau_\omega)$ and $\tilde{D}$ is the diagonal, $D$.

Because of these properties, the quantity $H$ is very useful at the research stage of prediction (the current stage) where the following problems are investigated:
- predictability of large earthquakes in principle;
- limits of predictability for $M \geq M_+$ earthquakes depending on the information used, e.g., earthquake catalogs.

The first problem can be interpreted as estimation of the significance level $\alpha$ of the relation $H = 0$, and the second as finding a reliable upper bounds on $H$. As a rule the difficulties here consist in the too short period of prediction monitoring and in the limited amount of data available for the estimation of $\lambda(dg)$.

Among all earthquake prediction methods available at present, the M8 algorithm (Keilis-Borok & *Kossobokov, 1990*) with its ongoing $M \geq 8.0$ worldwide prediction test (hereinafter M8.0+) seems to be the best adapted to statistical analysis because of a sufficiently long period of its monitoring and a large area of investigation. The M8 has been tested starting from 1985; since 1992 the test is in real time (*http://www.mitp.ru/en/predlist.html; Kossobokov et al., 1999; Kossobokov & Shebalin, 2003; Kossobokov, 2005*). In 1985-2009 about two dozen $M_+ \geq 8.0$ earthquakes occurred in the area of the M8 monitoring; more than half of these have been predicted. The M8.0+ monitoring space consists of a set of overlapping circles $B_R$ of radius $R = 668$ km (alarm unit) located along the Circum-Pacific and Alpine-Himalayan belts so as to cover the entire seismic zone (Fig. 1, the coordinates of the circles are listed in Table A, see Appendix). Despite the large area of the M8.0+ circles, a straightforward estimation of the target earthquake rates $\omega_i = \lambda(B_R^i | M_+) / \lambda(G | M_+)$ based on available catalogs is unreliable. Therefore earthquakes of lower magnitudes are commonly used for estimation of $\lambda(dg)$, which however does not remove completely the uncertainty of $\{\omega_i\}$ and generates additional problems.

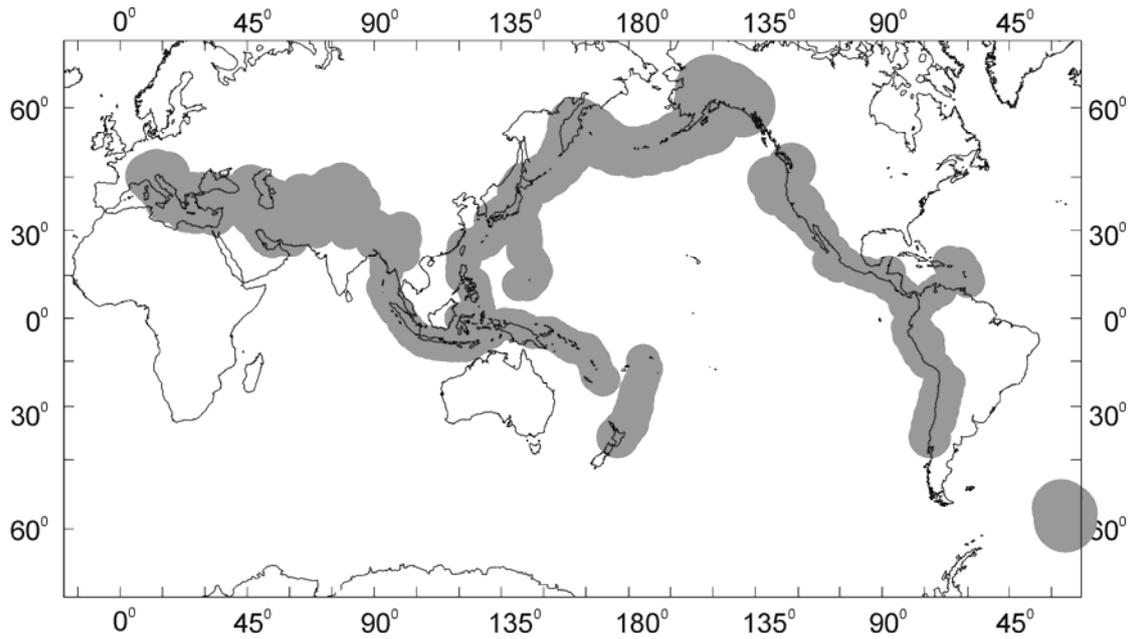

**Figure 1.** The M8.0+ test area: union of 262 overlapping circles of radius $R = 668$ km. Coordinates of the circle centers are listed in Appendix.

The M8 algorithm is a notable phenomenon in statistical seismology, and its prediction capability has been analyzed more than once by various researchers (see e.g., *Minster & Williams, 1998; Marzocchi et al., 2003*). The authors of the M8 algorithm publish the results of the monitoring on a regular basis, and in particular the estimates of $\tau_\omega$, which are based on the rates of smaller (usually $M \geq 4.0$) earthquakes. Marzocchi et al. (2003) asserted that these $\tau_\omega$ are significantly underestimated, and hence the prediction capability of the M8 is overestimated. Actually, this conclusion cannot be correct without consideration of the real M8 alarms.

Below we investigate theoretically the problem of significance of prediction results when $\omega(dg)$ is known inaccurately, and apply the theoretical results to the analysis of the M8 algorithm.

This work is closely connected with the investigations of the Collaboratory for the Study of Earthquake Predictability, CSEP (*Jordan, 2006; Schorlemmer et al., 2007; Zechar & Jordan, 2008*) focused on testing the earthquake prediction methods.

## 2. The theoretical aspect of prediction analysis

The rates of $M \geq M_+$ target earthquakes are not known exactly, and their empirical estimates are unreliable due to the very limited period of earthquake observation. That is why the earthquakes of lower magnitudes $M \geq M_-$ are commonly used for the estimation. The uncertainty of $\omega(dg)$ influences the estimation of the space-time alarm rate $\tau_\omega = \int_G \tau(g)\omega(dg)$, hence the estimation of the prediction significance. To analyze the problem we use the following error model for the measure $\omega(dg)$.

*2.1 The error model for $\omega(dg)$*

Divide the space $G$ into a set of nonintersecting areas $\{G_i, i = 1,...,k\}$, so that $\tau(g) = \tau_i$ for $g \in G_i$ and $\omega_i = \lambda(G_i / M_+) / \lambda(G / M_+)$. Suppose that the estimates $\{\hat{\omega}_i\}$ of $\{\omega_i\}$ are based on $M \geq M_+$ earthquakes, $N_\omega$ in number. The uncertainty of $\{\omega_i\}$ can be specified by a confidence zone of level $1 - \varepsilon$

$$\Omega_\varepsilon: \quad \sum_{i=1}^{k}(\hat{\omega}_i - \omega_i)^2 / \hat{\omega}_i < q_\varepsilon \tag{5}$$

with the appropriate threshold $q_\varepsilon$. The left part of (5) to be multiplied by $N_\omega$ is a statistic of the $\chi^2$ type with $f = k - 1$ degrees of freedom (*Rao, 1965*). The connection with the $\chi^2$ distribution becomes more legitimate if we make the following assumptions: the estimates $\{\hat{\omega}_i\}$ are unbiased, i.e., the expectation of $\hat{\omega}_i$ is $\omega_i$; the events used to estimate $\lambda(G_i | M_+)$ are weakly interdependent for different $G_i$; and $N_\omega$ is large enough. Then $N_\omega q_\varepsilon$ can be obtained as the $Q(1-\varepsilon, k-1)$ quantile of level $1 - \varepsilon$ for the $\chi_f^2$ distribution with $f = k - 1$, i.e.,

$$N_\omega q_\varepsilon = Q(1 - \varepsilon, k - 1). \tag{6}$$

*2.2 The confidence zone of random strategies*

The difficulty of using the $(n, \tau_{\hat{\omega}})$ plane for the estimation of space-time prediction results consists in the identification of random guess (or trivial) strategies. These strategies form a convex neighbourhood of the diagonal $n + \tau_{\hat{\omega}} = 1$ whose size depends both on the $\omega$ and $\hat{\omega}$ measures (*Molchan, 2010*). The union of such zones, $\hat{D}_\varepsilon$, over all $\omega$ from (5) can be considered as a confidence zone of all trivial strategies on the $(n, \tau_{\hat{\omega}})$ plane with confidence level $1 - \varepsilon$.

According to Molchan (2010), the zone $\hat{D}_\varepsilon$ under condition (5) is a convex polygon which contains the diagonal $n + \tau_{\hat{\omega}} = 1$ of the square $0 \leq n, \tau_{\hat{\omega}} \leq 1$. For applications it is important to consider the lower boundary of $\hat{D}_\varepsilon$, which is below the diagonal. Its vertices are located on the curve $n = f(\tau)$, where

$$f(\tau) = \begin{cases} 1 - \tau - \sqrt{q_\varepsilon \tau(1-\tau)}, & (1 + q_\varepsilon)\tau < 1 \\ 0, & (1 + q_\varepsilon)\tau > 1 \end{cases} \tag{7}$$

The zone $\hat{D}_\varepsilon$ can be slightly expanded by taking $f(\tau)$ as a new lower boundary. Then the expanded confidence zone for the trivial strategies

$$\hat{D}_\varepsilon^*: \quad f(\tau) \leq n \leq 1 - \tau, \quad 0 < \tau < 1, \tag{8}$$

will have a confidence level $\geq 1 - \varepsilon$ (Fig. 2).

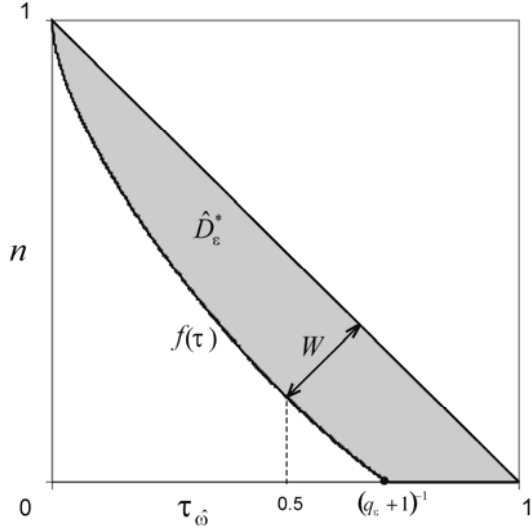

**Figure 2.** Confidence zone $\hat{D}_\varepsilon^*$ of level $1-\varepsilon$ for random strategies on the $(n, \tau_{\hat{\omega}})$ plane. The zone is specified for $\varepsilon = 0.01$, $k = 65$ (see (6)), and $N_\varepsilon = 238$ (the number of $Ms \geq 7.5$ events in the 1900-1984 Pacheco-Sykes catalog within the M8.0+ monitoring space).

*Notation*: $f(\tau)$ is the lower boundary of $\hat{D}_\varepsilon^*$ (see (7)); $W = h_\varepsilon/\sqrt{2}$ is width of the zone.

The zone $\hat{D}_\varepsilon^*$ is of interest because it does not depend on $\{\hat{\omega}_i\}$ and is specified by two parameters, $\varepsilon$ and $N_\omega$. For the sake of simplicity $\hat{D}_\varepsilon^*$ can be used in probabilistic prediction (*Zechar & Jordan, 2008*) and, as we will show, for a restriction on $N_\omega$. In fact, one important parameter of the $\hat{D}_\varepsilon^*$ zone is its width

$$h_\varepsilon = \max_{\hat{D}_\varepsilon^*}(1 - n - \tau_{\hat{\omega}}) = \max_\tau \sqrt{q_\varepsilon \tau(1-\tau)} = \sqrt{q_\varepsilon}/2, \qquad (9)$$

which is measured in units of the distance from the origin (0,0) to the diagonal $n + \tau_{\hat{\omega}} = 1$. The value of $h_\varepsilon$ is an upper bound on the apparent predictive capability of any trivial strategy on the $(n, \tau_{\hat{\omega}})$ plane. It is therefore natural to use the restriction

$$\sqrt{q_\varepsilon}/2 \leq h_0, \qquad (10)$$

where $h_0$ is a control parameter.

By (6,10) we have

$$N_\omega \geq Q(1-\varepsilon, k-1)/4h_0^2. \qquad (11)$$

The M8 algorithm uses overlapping circles $G_i = B_R(i)$. In order to be able to use the estimate (11), we find $k$ as follows: $k = areaG/areaB_R \approx 65$, i.e., $k$ is the number of nonintersecting circles $B_R$ whose total area is the same as for the prediction space $G$.

Substituting in (11) the values $k = 65$ and $\varepsilon = 0.01$, we get the table:

| $h_0$ | 5% | 5.5% |
|---|---|---|
| $N_\omega$ | 9322 | 7704 |

Note that the 1977-2004 CMT catalog to be used below contains 8508 $Mw \geq 5.5$ events and 2843 $Mw \geq 6.0$ events within the M8.0+ monitoring space. The $Mw \geq 5.5$ data guarantee that $h_\varepsilon \leq 5.2\%$, while with $Mw \geq 6.0$ the value of $h_\varepsilon$ becomes too great, $h_\varepsilon > 9\%$. Thus, we suppose to be reasonable $M_- \leq 5.5$ for $Mw$. Similarly, we have $M_- \leq 5.0$ for $Ms$ (NEIC, 1969-2004) and $M_- \leq 5.5$ for $mb$ (NEIC, 1963-2004). Note that a "correct" $\omega(dg)$ estimate by Marzocchi et al. (2003) is based on the $Ms \geq 7.5$ events from the 1900-1984 Pacheco-Sykes (1992) catalog. In that case we have $N_\omega = 238$, i.e., on the average less than 4 events per one of a 65 circles. Consequently, $h_\varepsilon$ is inadmissibly high here, $h_\varepsilon > 30\%$ (see also Fig.2).

*2.3. Significance level of prediction results*

The prediction results for a period $T^+$ are described by $(\nu; N; \hat{\tau}(g), g \in G)$, where $\nu$ is the number of failures-to-predict, $N$ the number of target events for the time $T^+$, and $\hat{\tau}(g)$ the actual relative alarm time at the point $g$.

Statistical significance of the result is estimated relative to an assumed model of target events. The nature of large target events is far from being understood (see, e.g., *Romanowicz, 1993*). The conventional null approximation is the Poisson model which is homogeneous in time and inhomogeneous in space with the rate measure $\lambda(dg \mid M_+)$ (the $H_0$ hypothesis).

Given $(N, \hat{\tau}(g))$, the conditional probability to have $\nu$ or less failures-to-predict under $H_0$ determines the observed significance of the prediction results, $\alpha$. This quantity can be found from the binomial distribution with parameter $N$ (the number of trials) and the probability of a success (here, random success),

$$\tilde{\tau} = \int_G \hat{\tau}(g)\omega(dg), \tag{12}$$

as follows:

$$\alpha = \sum_{i=0}^{\nu} \binom{N}{i} \cdot \tilde{\tau}^{N-i}(1-\tilde{\tau})^i = \binom{N}{\nu} \cdot \int_0^{\tilde{\tau}} (1-x)^\nu dx^{N-\nu}, \tag{13}$$

where the $\binom{N}{i}$ are the binomial coefficients.

To estimate $\alpha$ in the conditions (5), note that the right-hand side of (13) is an increasing function of $\tilde{\tau}$. (We recall that the higher the value of $\alpha$, the greater degree of randomness is exhibited by the results.) It follows that we can derive an upper estimate of $\alpha$ from an upper estimate of $\tilde{\tau}$.

When the space is discretized, $G = \bigcup G_i$, with nonintersecting $G_i$, the admissible variations in $\omega_i = \omega(G_i)$ are given by (5) and $\tilde{\tau} = \sum \hat{\tau}_i \omega_i$. Hence, to find $\max \tilde{\tau}$ we must take into account (5) and the relation $\sum_{i=1}^k \omega_i = 1$. This problem is easily solved by using Lagrange multipliers. The result is

$$\max_{\Omega_\varepsilon} \tilde{\tau} = \hat{\tau} + \sqrt{q_\varepsilon} \sigma_\tau, \tag{14}$$

where

$$\hat{\tau} = \int_G \hat{\tau}(g)\hat{\omega}(dg), \tag{15}$$

$$\sigma_\tau^2 = \int_G \hat{\tau}^2(g)\hat{\omega}(dg) - \hat{\tau}^2 . \qquad (16)$$

Here, $q_\varepsilon$ is given by (6) and $\hat{\omega}$ is an unbiased estimate of $\omega$ based on the total number $N_\omega$ of the $M \geq M_-$ events.

Since $0 \leq \hat{\tau}(g) \leq 1$, we have $\sigma_\tau \leq 0.5$. Consequently, the maximum disturbance of $\hat{\tau}$ due to the uncertainty in $\omega(dg)$ does not exceed $h_\varepsilon$ (see (14) and (9)). In turn the restriction $h_\varepsilon \leq h_0$ depends on the sample size $N_\omega$ (see (11)). The substitution $\tilde{\tau} = \hat{\tau} + \sqrt{q_\varepsilon}\sigma_\tau$ in (13) yields an upper estimate of $\alpha$.

It remains to recall the assumptions that underlie this bound:
- the $\omega(G_i)$ estimates are non-biased;
- the $M \geq M_-$ events in different $G_i$ used for $\{\hat{\omega}(G_i)\}$ are weakly interdependent;
- the numbers of $M \geq M_-$ events in $\{G_i\}$ used to estimate $\{\omega(G_i)\}$ are not small (greater than 10 in our calculations).

## 3. Analysis of prediction results: the M8 algorithm
### 3.1. The $H_b$ and $H_{GR}$ hypothesis

The estimation of alarm volume $\tau_\omega$ is the central problem in the analysis of significance for prediction results. In M8.0+ this quantity depends on the measure $\omega(dg)$ over the circles $B_R$ of a large radius $R=668$ km, $\omega_i = \omega(B_R(i))$. Straightforward estimates of $\omega_i$ based on large ($M_+ = 8.0$) events are very unstable. A possible way out of this difficulty is to use $M > M_-$ events ($M_- < M_+$), because they are much more numerous.

The following arguments can be advanced to support this decision for the M8 algorithm:
- Since the circles $B_R$ are large, this allows us to disregard the specific distribution of smaller events that deviate from the main faults, which are as a rule responsible for large earthquakes;
- The circles are large compared with the rupture lengths of the target events. This provides a necessary condition for the recurrence curve to be linear in the range $(M_-, M_+)$ (*Molchan et al., 1997*). Assuming the Gutenberg-Richter relation (G-R) to be linear on $(M_-, M_+)$ (*the $H_{GR}$ hypothesis*), we can find the rate of large events from that of smaller ones as follows:

$$\lambda(B_R(i) \mid M_+) = C_i \lambda(B_R(i) \mid M_-), \qquad (17)$$

where

$$C_i = 10^{-b_i(M_+ - M_-)}, \qquad (18)$$

and $b_i$ is the b-value in the circle $B_R(i)$ for the range $(M_-, M_+)$;
- The hypothesis of the *b*-value universality is considered as admissible by some researchers (see, e.g., *Kagan, 1999, 2009b*). Using the 1982-1997 Harvard catalog Kagan (1999) concludes that the *b*-value is universal for events within 500 km depth and with magnitudes $5.5 \leq Mw \leq 8.1$. The hypothesis that the *b*-value is independent of region for the $M_- \leq M \leq M_+$ events will be referred to as *the $H_b$ hypothesis*. Under $H_b$ the coefficients $C_i$ in (17) are independent of $i$. Consequently, the normalized rates $\omega_i$ of the $M \geq M_-$ and $M \geq M_+$ events are identical. Note that the use of $\omega_i$ estimates based on $M \geq M_-$ instead of $M \geq M_+$ automatically implies that the $H_b$ hypothesis is true.

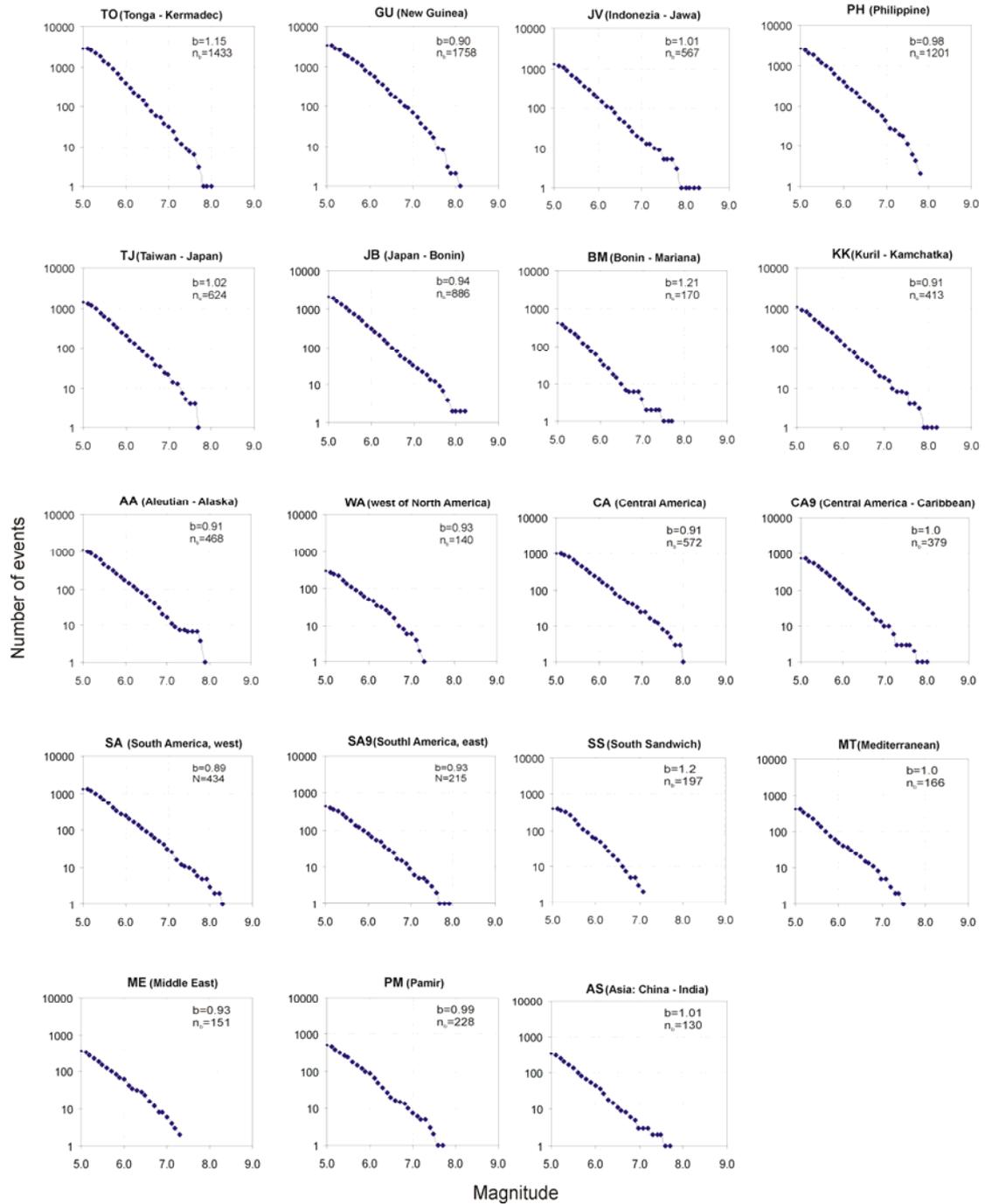

**Figure 3.** Cumulative frequency-magnitude relation for the sub-regions of the M8.0+ test area (see Appendix for abbreviation). The data are from the 1977-2008 CMT catalog.
*Notation*: $b$ is the likelihood estimate of the $b$-value for $Mw \geq 5.5$; $n_b$ is the number of $Mw \geq 5.5$ events.

Unfortunately, the hypothesis of a universal *b*-value is doubtful (see, e.g., *Molchan et al., 1997*), while small variations in the *b*-value can affect the significance estimate. Figure 3 shows a sequence of frequency-magnitude relations for 19 sub-regions in the M8.0+ prediction space. The sets of M8.0+ circles that make the sub-regions are listed in Table A, see Appendix. Figure 3 is based on the 1977-2004 CMT catalog.

To interpret the linearity of the empirical relation $\lg N(m)$ vs. $m$ where $N(m)$ is number of $M \geq m$ events, we remind the following. By the Jensen's inequality, one has

$$E \lg N(m) < \lg E N(m) = a - bm.$$

For large $N(m)$ the quantities $\lg N(m)$, $E \lg N(m)$, and $\lg E N(m)$ are close with each other, but it is not the case otherwise. Therefore, even under $H_{GR}$ hypothesis we can observe a downward bend of the G-R curve when $N(m) \leq 10$. It follows from Fig. 3 that the frequency-magnitude relations in the range $Mw : 5.5 - 8.0$ can be assumed to be linear for nearly all sub-regions. Note that a similar conclusion for *Ms* data is not so confident.

Figure 3 contains the *b*-value estimates obtained by the likelihood method for the magnitude range $Mw \geq M_- = 5.5$ (see e.g. *Molchan et al., 1997*). The number of events in the sub-regions varies from 130 to 1758, which guarantees the accuracy $\sigma_b \leq 0.1$ for the *b*-value estimate. We have $b = 0.9 - 1.0$ for 16 sub-regions. The remaining three (Tonga-Kermadec trenches (TO), Bonin-Mariana trenches (BM), and South Sandwich islands (SS)) have their *b*-values above the normal level, more specifically, $b \approx 1.2$.

The following effect arises from the *b*-value being inhomogeneous. Suppose the $\{\omega_i\}$ are based on the $M \geq M_-$ seismicity. In virtue of (17) the weight ratio $\omega_i / \omega_j$ is reduced by a factor of $\gamma = 10^{(b_i - b_j)(M_+ - M_-)}$, if the $\omega_i$ are calculated using the G-R relation (see (17)). Putting $b_i = 1.2$, $b_j = 0.9 - 1.0$, and $M_+ - M_- = 8.0 - 5.5 = 2.5$, we have $\gamma = 3 - 5.5$. The estimates of $\tau_\omega$ may be sensitive to such changes in $\{\omega_i\}$ (see below).

The empirical estimate of the measure of the set $A$, $\omega(A)$ based on $M \geq M_-$ is given by

$$\hat{\omega}(A) = n(A | M_-) / n(G | M_-), \quad (19)$$

where $n(A | M_-)$ is the number of $M \geq M_-$ events in $A$ for the period $T^-$.

The estimate of $\omega(A)$ based on the G-R relation, i.e., under $H_{GR}$, is derived differently. Denote by $\{G_j\}$ the 19 sub-regions of $G$ mentioned above and by $\{b_j\}$ the set of their respective *b*-values. For $A \subset G$ we have the estimate

$$\hat{\omega}(A) = n_{GR}(A | M_-) / n_{GR}(G | M_-), \quad (20)$$

where

$$n_{GR}(A / M_-) = \sum_i [g_i \in A] \cdot 10^{-b_{k(i)}(M_+ - M_-)}. \quad (21)$$

Here $[\circ]$ is the (1,0) logical function, $g_i$ is $M \geq M_-$ event in the volume $G \times T^-$, and $k(i) = j$ if $G_j$ contains the circle $B_R$, whose centre is the nearest to $g_i$.

*3.2. The data.*
Our earthquake sources include:

−   the National Earthquake Information Center Global Hypocenters Database, NEIC (*GHDB, 1989*) routinely updated through 2004 from NEIC Preliminary Determination of Epicenters data (PDE) (http://earthquake.usgs.gov/research/data/). The duplicates have been removed using the automatic procedure by Shebalin (1992). The database contains earthquakes with magnitude *mb* for the period 1963-2004 and *Ms* for 1969-2004;

−   the Centroid Moment Tensor catalog, CMT (*Ekstrom et al., 2005*) (since 2007 the Global Centroid Moment Tensor catalog), *Mw* magnitude, 1977-2004;

−   the Global Pacheco-Sykes Catalog (*Pacheco & Sykes, 1992*), *Ms* magnitude, 1900-1984.

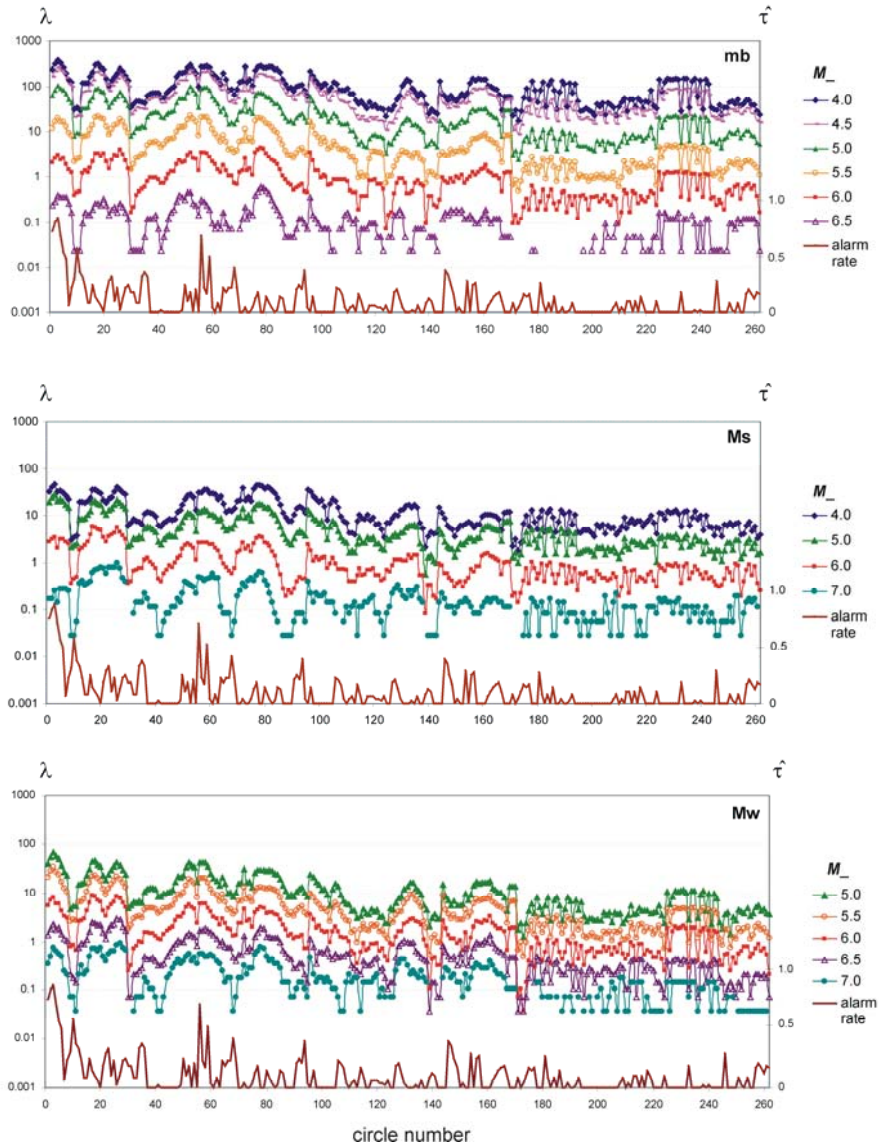

**Figure 4.** Alarm rate $\hat{\tau}(B_R)$ ( right-hand axis) for 262 circles of the M8.0+ test and the annual rate of $M \geq M_-$ events, $\lambda(B_R)$, for the same circles (left-hand axis); the rate $\lambda(B_R)$ depends on $M_-$ and magnitude type: *mb*, NEIC, 1963-2004 (top); *Ms*, NEIC, 1969-2004 (middle); and *Mw*, CMT, 1977-2004 (bottom). The alarm rate is given for the period 1985-2009.

Figure 4 represents the rates of $M \geq M_-$ events in 262 M8.0+ circles with different thresholds $M_-$ for the NEIC and CMT catalogs. The vertical $\lambda$-axis is on a log scale by analogy with the Gutenberg-Richter relation, so that the nearly constant distance between adjacent plots supports the hypothesis of an exponential distribution for earthquake magnitudes in the circles. Figure 4 shows that the completeness level in the circles varies between 4.5 and 5.0 for $mb$, is around 5.0 for $Ms$, and around 5.5 for $Mw$. In addition, there are large differences in the overall seismicity level from circle to circle (by factors of 50 or 60).

Combining the estimates of completeness for $Mw$, $Ms$, $mb$ with the upper bounds $M_-$ in Section 2.2, we come to the preferable choice of $M_-$: $M_- = 5.5$ for $Mw$, and $M_- = 5.0$ for $Ms$ and $mb$.

### 3.3. The $\tau_\omega$ estimates

We recall the main parameters which are required to obtain $\hat{\tau}$:
- magnitude type, viz. $mb$, $Ms$, $Mw$;
- magnitude threshold $M_-$. The $M \geq M_-$ events are used to estimate $\omega(g)$, below $M_- = 4.5 - 7.0$;
- the past time period $T^-$ which contains the $M \geq M_-$ events. Two variants of $T^-$ are considered. The first, $T_1^-$, terminates before the beginning of the M8.0+ monitoring, i.e., by January 1, 1985; and the second, $T_2^-$, by December 26, 2004, i.e., before the Sumatra-Andaman mega-earthquake, which has noticeably affected the seismic background in the M8 prediction space. Both intervals have a common initial time which depends on magnitude type: 1963 for $mb$, 1969 for $Ms$, and 1977 for $Mw$. We use these two intervals for the following reasons. The extended time interval gives more data, which improves the stability of $\hat{\omega}(dg)$. However, the earthquakes that have occurred during the monitoring period (after January 1, 1985) and the target events are not independent, and this can affect the estimate of $\tau_\omega$. In particular, the aftershocks, which have not been eliminated from the catalogs, can increase the target event rate estimate within the correct alarm zone and thus reduce the significance of the prediction results. But the situation is not so simple, and on the whole the effect can disappear because the local weights $\hat{\omega}(dg)$ are normalized and included in $\hat{\tau}$ as $\hat{\tau}(g)\hat{\omega}(dg)$;
- the time period of monitoring $T^+$. There are two variants of $T^+$: the whole M8.0+ testing time, $T_1^+ = 1985 - 2009$, and the forward-prediction time, $T_2^+ = 1992 - 2009$;
- the type of $\omega(g)$ estimate, i.e., (19) or (20). The first is based formally on $M \geq M_-$ events only, the second also uses the $H_{GR}$ hypothesis.

Table 1 summarizes the $\hat{\tau}$ estimates. For the sake of simplicity the $\hat{\tau}$ are shown for $T_2^-$ only; the $\hat{\tau}$ for $T_1^-$ are slightly smaller (as a rule the difference is less than 0.01). The extreme numbers of events, $N_\omega$, used for the $\hat{\tau}$ differ by a factor of two hundred. Nevertheless, the values of $\hat{\tau}$ vary within 0.3-0.4. The greatest values of $\hat{\tau}$ for different time periods and different magnitude types occur at $M_- = 5.0$ (for $mb$, $Ms$) and at $M_- = 5.5$ (for $Mw$).

**Table 1.** $\hat{\tau} \cdot 100$ (%) for M8.0+ alarm depending on magnitude type, threshold $M_-$, and monitoring period. All $\omega$- estimates are based on data in the time interval $T_2^-$; $N_\omega$ varies from 863 ($Ms \geq 6.5$) to 139588 ($mb \geq 4.5$); the $\hat{\tau}$ estimates in brackets are based on small numbers of events $N_\omega = 227 - 314$.

| Magnitude threshold M_ | $mb$ | | $M_S$ | | $M_W$ | |
|---|---|---|---|---|---|---|
| | 1985-2009 | 1992-2009 | 1985-2009 | 1992-2009 | 1985-2009 | 1992-2009 |
| **4.5** | 37.9 | 35.8 | 37.3 | 34.8 | - | - |
| **5.0** | 39.6 | 38.0 | 38.7 | 36.2 | 40.7 | 38.3 |
| **5.5** | 37.9 | 36.4 | 37.5 | 35.0 | 41.3 | 38.8 |
| **6.0** | 36.1 | 34.3 | 36.7 | 34.3 | 38.7 | 36.1 |
| **6.5** | *(38.6)* | *(37.2)* | 35.0 | 32.8 | 37.6 | 34.7 |
| **7.0** | - | - | *(35.2)* | *(33.5)* | *(38.3)* | *(36.0)* |
| $H_{GR}$ hypothesis | | | | | | |
| **5.0** | - | - | 35.7 | 33.1 | - | - |
| **5.5** | - | - | 34.2 | 31.7 | 35.4 | 32.5 |
| **6.0** | - | - | 32.2 | 29.6 | 35.4 | 32.3 |

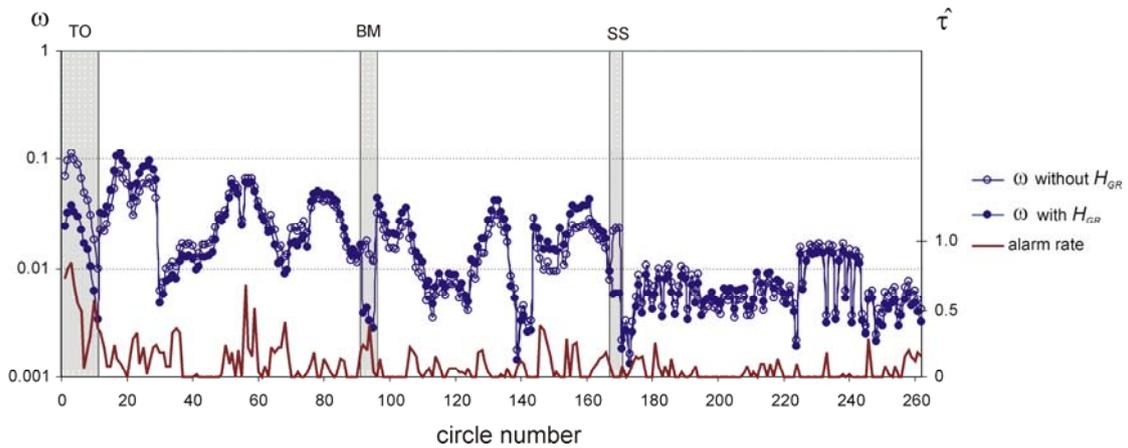

**Figure 5.** Alarm rate $\hat{\tau}(B_R)$ ( right-hand axis) and normalized rate of $Mw \geq 5.5$ events, $\omega(B_R)$, (left-hand axis) for 262 circles of the M8.0+ test. Two upper curves correspond to $\omega(B_R)$ obtained under the $H_{GR}$ hypothesis (filled circles) and when $H_{GR}$ is not used (open circles). The intervals TO, BM, SS correspond to sub-regions (see Appendix) with a non-standard b-value, $b \approx 1.2$.

In particular, for $M_- = 5.5$ ($Mw$) and $T_1^+$ we have $\hat{\tau} \cdot 100 \approx 35\%$ under the $H_{GR}$ hypothesis and $\hat{\tau} \cdot 100 \approx 41\%$ when $H_{GR}$ is not used. The difference in the estimates is related to the *b*-value effect discussed above. It can be observed by comparing the estimates $\hat{\omega}(B_R(i))$ derived for $Mw$ with and without regional *b*-values (Fig. 5). As was to be expected, the zones with higher *b*-values ($b = 1.2$), i.e., TO, BM and SS, exhibit considerable differences in $\omega_i$. Figure 5 also shows the alarm rate $\tau_i$ in circles $B_R(i)$. One can see that the alarm rates in TO and BM are substantially greater than the mean level. Hence we have here a resonance of two effects, and, as a result, the observed variations in $\hat{\tau}$. Simultaneously we get an example of how small events can inflate the estimate $\hat{\tau}$.

As a supplement to Table 1, we estimated $\hat{\tau}$ using the $mb \geq 4.0$ events from the NEIC and the $Ms \geq 7.5$ events from the Pacheco-Sykes catalog. According to Marzocchi et al. (2003), it is only the second estimate which can be correct, and its value must be substantially greater than that of the first. As a matter of fact, the two estimates are nearly identical: for the period 1985-2009 we have $\hat{\tau}(mb \geq 4.0) \cdot 100 = 36.2\%$ and $\hat{\tau}(Ms \geq 7.5) \cdot 100 = 36.3\%$. Similarly to the preceding example, we conclude that the differences in the $\hat{\tau}$ estimates cannot be correctly interpreted, unless the respective alarms are taken into account.

What is much more important is that use of the both estimates in the analysis of M8 is questionable: the former because the $H_b$ hypothesis requires substantiation for the magnitude interval $4.0 - 8.0$, and the latter in view of the scarcity of data, $N_\omega = 238$. In other words, the choice of $M_-$ faces two opposite tendencies: with increasing $M_-$ the $H_{GR}$ hypothesis about the frequency-magnitude relation in $(M_-, M_+)$ being linear becomes more likely; but at the same time the amount of available data $N_\omega$ decreases, thereby making the uncertainty of $\hat{\tau}_\omega$ larger.

### 3.4. The significance of M8.0+ results, $\alpha$

*Parameters.* To restrict the number of options for the estimation of $\alpha$ we fix the following priorities:

- $H_{GR}$ is considered as the main hypothesis, since it is more flexible than $H_b$ and is in general agreement with the data in the sub-regions; the estimates of $\hat{\tau}$ (Table 1) obtained without $H_{GR}$ are auxiliary and can be used for comparison;

- Of all magnitude scales, $Mw$ is the most consistent with the $H_{GR}$ hypothesis in $(M_-, M_+)$; $mb$ is unsuitable for $H_{GR}$ because of its saturation at large $M$;

- $M_- = 5.0$ and 5.5 are considered as the preferred values of $M_-$. The reason for this is as follows: the quantity $M_-$ is bounded from below by the magnitude of complete reporting and from above by a threshold that follows from the restrictions on $N_\omega$ in (11). As it is shown above, the choice $M_- = 5.5$ for $Mw$ and $M_- = 5.0$ for $Ms$ and $mb$ magnitudes seems to be optimal. Moreover, the greatest values of $\hat{\tau}$ in Table 1 are reached for $M_- = 5.0; 5.5$.

In order to calculate the perturbation of $\hat{\tau}$ due to the errors in $\omega(dg)$ we use (14), namely, $\delta\hat{\tau} = \sqrt{Q(1-\varepsilon, k-1)/N_\omega} \cdot \sigma_\tau$ with $k = 65$ and $\varepsilon = 0.01$. Then we have $Q = 93.2$. The values of $N_\omega$ and $\sigma_\tau$ for the period $T_2^-$ are given in Tables 2 and 3, respectively. The quantity $\sigma_\tau$ is very stable: $\sigma_\tau \approx 0.25 - 0.30$ for different types of magnitude ($mb$, $Ms$, $Mw$), thresholds $M_-$ (5.0, 5.5), periods of monitoring ($T_1^+$, $T_2^+$), and the hypotheses ($H_b$, $H_{GR}$). Consequently $\delta\hat{\tau} \approx 2.5 - 4\%$.

The ultimate values of $\tilde{\tau} = \hat{\tau} + \delta\hat{\tau}$ under the $H_{GR}$ hypothesis are given by Table 4. We have $\tilde{\tau} = 0.35 - 0.38$, $\hat{\tau} = 0.32 - 0.36$ for the two monitoring periods and two magnitude scales, $Ms$ and $Mw$.

**Table 2.** Number of events $N_\omega$ for the period $T_2^-$ depending on magnitude type and threshold $M_-$.

| Magnitude threshold M_ | $mb$ 1963-2004 | $M_S$ 1969-2004 | $M_W$ 1977-2004 |
|---|---|---|---|
| **5.0** | 51520 | 10792 | 18078 |
| **5.5** | 11947 | 5033 | 8508 |

**Table 3.** $\hat{\sigma}_\tau$ for M8.0+ alarm depending on magnitude type, threshold $M_-$, and monitoring period. The $\omega$- estimates are the same as in Table 1.

| Magnitude threshold M_ | $mb$ | | $M_S$ | | $M_W$ | |
|---|---|---|---|---|---|---|
| | 1985-2009 | 1992-2009 | 1985-2009 | 1992-2009 | 1985-2009 | 1992-2009 |
| **5.0** | 0.29 | 0.32 | 0.30 | 0.32 | - | - |
| **5.5** | 0.29 | 0.31 | 0.28 | 0.30 | 0.31 | 0.33 |
| $H_{GR}$ hypothesis | | | | | | |
| **5.0** | - | - | 0.24 | 0.28 | - | - |
| **5.5** | - | - | 0.24 | 0.28 | 0.25 | 0.28 |

**Table 4.** Upper estimate of $\tau$, $\tilde{\tau} \cdot 100\%$, under the $H_{GR}$ hypothesis.

| Magnitude threshold M_ | $M_S$ | | $M_W$ | |
|---|---|---|---|---|
| | 1985-2009 | 1992-2009 | 1985-2009 | 1992-2009 |
| **5.0** | 37.9 | 35.7 | - | - |
| **5.5** | 37.5 | 35.5 | 38.0 | 35.4 |

*Target events.* It remains to find the number of target events for the monitoring period. This is the most sensitive point for calculating the significance of the M8. The M8 algorithm uses as the working magnitude the greatest of those provided for earthquake in the NEIC catalog. The uncertainty of the working magnitude appears to be $\approx 0.3$, while the target events are counted exactly following the prescribed threshold $M_+ = 8.0$. Table 5 gives the list of $M \geq 8.0$ events from the NEIC catalog occurred in 1985-2009 in the M8.0+ prediction space. The table does not contain the March 12, 1995 Iturup, $Ms = Mw = 7.9$ earthquake. It is considered as a success on the M8 web-site because the reported magnitude was $Mmax$(NEIC)=8.0 up to 1997.

**Table 5.** List of large, $M_{max} \geq 8.0$, earthquakes from the NEIC catalog, which occurred in the area of M8.0+ monitoring in 1985-2009.

| Region | Date | Lat | Lon | Magnitude | | | | M8.0+ result |
|---|---|---|---|---|---|---|---|---|
| | | | | mb | Ms | m3 | m4 | |
| Mexico | 1985.09.09 | 18.19 | -102.53 | 6.8 | 8.1 | 7.9MsBRK | 7.9MsPAS | Yes |
| Kermadec | 1986.10.20 | -28.12 | -176.37 | 6.6 | 8.1 | 8.3MsBRK | - | Yes |
| Guam | 1993.08.08 | 12.98 | 144.80 | 7.1 | 8.0 | 8.2MsBRK | 7.8MwHRV | Yes |
| Kurils | 1994.10.04 | 43.77 | 147.32 | 7.3 | 8.1 | 7.9MsBRK | 8.3MwHRV | Yes |
| Samoa | 1995.04.07 | -15.20 | -173.53 | 6.8 | 8.0 | 8.1MsBRK | 7.4MwGS | Yes |
| N. Guinea | 1996.02.17 | -0.89 | 136.95 | 6.5 | 8.1 | 8.2MwHRV | 7.7MeGS | Yes |
| Sumatra | 2000.06.04 | -4.72 | 102.09 | 6.8 | 8.0 | 8.3MeGS | 7.9MwHRV | Yes |
| Indonesia | 2000.11.16 | -3.98 | 152.17 | 6.0 | 8.2 | 8.2MsBRK | 8.0MwHRV | No |
| Indonesia | 2000.11.17 | -5.50 | 151.78 | 6.2 | 8.0 | 8.2MsBRK | 7.8MwHRV | No |
| Peru | 2001.06.23 | -16.26 | -73.64 | 6.7 | 8.2 | 8.4MwHRV | 8.3MwGS | No |
| China | 2001.11.14 | 35.95 | 90.54 | 6.1 | 8.0 | 7.8MwHRV | 7.5MwGS | Yes |
| Alaska | 2002.11.03 | 63.65 | -147.44 | 7.0 | 8.5 | 7.9MwHRV | 8.1MeGS | No |
| Hokkaido | 2003.09.25 | 41.81 | 143.91 | 6.9 | 8.1 | 8.3MwHRV | 8.1MwGS | No |
| Sumatra | 2004.12.26 | 3.29 | 95.98 | 7.0 | 8.8 | 9.0MwHRv | 8.2MwGS | (*) |
| Sumatra | 2005.03.28 | 2.09 | 97.11 | 7.2 | 8.4 | 8.6MwHRV | 8.1MwGS | No |
| Tonga | 2006.05.03 | -20.19 | -174.12 | 7.2 | 7.8 | 8.0MwHRV | 7.9MwGS | Yes |
| Kurils | 2006.11.15 | 46.59 | 153.27 | 6.5 | 7.8 | 8.3MwHRV | 7.9MwGS | No |
| Kurils | 2007.01.13 | 46.24 | 154.52 | 7.3 | 8.2 | 8.1MwGCMT | 7.9MwGS | No |
| Solomon | 2007.04.01 | -8.47 | 157.04 | 6.8 | 7.9 | 8.1MwGCMT | 7.7MeGS | Yes |
| Peru | 2007.08.15 | -13.39 | -76.60 | 6.7 | 7.9 | 8.0MwGCMT | 7.7MeGS | No |
| Sumatra | 2007.09.12 | -4.44 | 101.37 | 6.9 | 8.5 | 8.5MwGCMT | 8.2MeGS | Yes |
| Sumatra | 2007.09.12 | -2.63 | 100.84 | 6.6 | 8.1 | 7.9MwGCMT | 7.8MeGS | Yes |
| China | 2008.05.12 | 31.00 | 103.32 | 6.9 | 8.1 | 7.9MwUCMT | 7.9MwGCMT | No |
| Samoa | 2009.09.12 | -15.49 | -172.09 | 7.1 | 8.1 | 8.1MwGCMT | 8.0MwUCMT | Yes |

(*) The 26 December, 2004, Sumatra mega-earthquake does not belong to M8.0+ class (see text for details).

Another difficulty is that the upper magnitude threshold of target events is not specified. An analysis of the 2004 Sumatra-Andaman mega-earthquake showed it to be outside the $M 8.0+$ range (*Kossobokov et al., 2009*; *Romashkova, 2009*). This is natural, because the M8 algorithm is adjusted based on similarity considerations. This is in particular applicable to the alarm area unit $B_R$, which for $M_+ = 9.0$ would be comparable with a hemisphere of the Earth. It is for the same reason that the authors of M8 use different parameters to predict M7.5+ and M8.0+ earthquakes. It follows that the range of the M8.0+ events under prediction should be restricted from above, say $M < 8.5$ by analogy with the M7.5+ test. The narrow range of target events and the relatively large magnitude uncertainty considerably complicate the counting of target events. Below we specify M8.0+ range in two ways: $8.0 \leq M < 8.5$ and $8.0 \leq M < 8.7$.

*The $\alpha$-estimates*. Now we are ready to estimate the observed significance $\alpha$ for prediction results in the M8.0+ experiment. Using $\hat{\tau}$ (Table 1), $\tilde{\tau}$ (Table 4), and (13), we get a point estimate $\hat{\alpha}$, which is based on $\hat{\tau}$, and an upper estimate $\tilde{\alpha}$, which is based on $\tilde{\tau}$ (see Table 6). Table 6 shows $\hat{\alpha}$, $\tilde{\alpha}$ for the two periods of monitoring, for the two ranges of target events, for $Mw \geq M_-$ with $M_- = 5.5$, and for the $H_{GR}$ hypothesis.

**Table 6.** Significance estimates for prediction results of the M8.0+ experiment. *Notation:* $\hat{\alpha}$ and $\tilde{\alpha}$ are point and upper estimates of the significance respectively; $N$ is the number of target events; $\nu$ is the number of failures-to-predict; $\tilde{H}$ is a conservative estimate of the M8.0+ prediction capability.

| Period | Target range | $(N-\nu)/N$ | $\alpha \cdot 100$ (%) | $\tilde{\alpha} \cdot 100$ (%) | $1-\hat{n}-\tilde{\tau} = \tilde{H} \cdot 100$ (%) |
|---|---|---|---|---|---|
| 1985-2009 | $8.0 \leq M < 8.7$ | 13/23 | 3.1 | 5.5 | 19 |
| | $8.0 \leq M < 8.5$ | 12/20 | 2.2 | 3.8 | 22 |
| 1992-2009 | $8.0 \leq M < 8.7$ | 11/21 | 4.7 | 8.3 | 17 |
| | $8.0 \leq M < 8.5$ | 10/18 | 3.7 | 6.4 | 20 |

All point estimates $\hat{\alpha}$ (including those for $Ms \geq M_- = 5.5$ and $Ms \geq M_- = 5.0$) vary within the range 2-5%, supporting the assertion that the M8 algorithm is non-trivial. Comparison of $\hat{\alpha}$ for the two target magnitude ranges shows that $\hat{\alpha}$ is smaller for the range $8.0 \leq M < 8.5$. In this case $\hat{\alpha} = 2.2\%$ for $T_1^+ = 1985 - 2009$ and $\hat{\alpha} = 3.7\%$ for $T_2^+ = 1992 - 2009$. As for the upper estimates $\tilde{\alpha}$, we have the nominal significance $\tilde{\alpha} < 5\%$ for the case ($8.0 \leq M < 8.5$, $T_1^+$) only, while for ($8.0 \leq M < 8.5$, $T_2^+$) it is $\tilde{\alpha} = 6.4\%$. However $\tilde{\alpha}$ is unstable because the numbers $(\nu, N)$ are small. Indeed, a next $8.0 \leq M < 8.5$ event in M8.0+ area will have resulted in $(\nu, N+1)$ to replace $(\nu, N)$ in the case of successful prediction, or else it will be $(\nu+1, N+1)$ in the case of a failure-to-predict. Then for the period $T_2^+$ we shall have: $\tilde{\alpha} = 3.8\%$ or $\tilde{\alpha} = 9.4\%$, respectively, instead of the observed $\tilde{\alpha} = 6.4\%$.

Table 6 also contains the score: the number of successes $(N-\nu)$ to the number of target events $(N)$, and a lower estimate of $H$: $\tilde{H} = 1 - \hat{n} - \tilde{\tau}$, where $\hat{n} = \nu/N$. The value $\tilde{H} \approx 20\%$ can therefore be considered as a conservative estimate of prediction capability for the M8 algorithm in prediction of $8.0 \leq M < 8.5$ events.

## 4. Conclusions and discussion

We have tried to provide a constructive answer to the question how to use correctly the inaccurate measure of target events $\omega(dg)$ to estimate the significance of prediction results and to find a confidence zone of trivial strategies on the $(n, \tau_{\hat{\omega}})$ diagram.

In application to the results of M8.0+ prediction our analysis can be summed up as follows. The magnitude range of the target events is in need of additional specification. The interval $8.0 \leq M < 8.5$ seems to be most appropriate restriction for M8.0+. In this case there are 12 successes of 20 target events for the whole period of monitoring, 1985-2009. Then the point estimate of significance is $\hat{\alpha} = 2.2\%$ and the upper estimate is $\tilde{\alpha} = 3.8\%$. However, for the period of advance prediction, 1992-2009, the score is 10 successes of 18, $\hat{\alpha} = 3.7\%$, and $\tilde{\alpha} = 6.4\%$. On the whole, these results support the non-triviality of M8 as a predictor and at the same time indicate possible instability of the $\alpha$-estimates. The prediction capability of M8.0+, i.e. the rate of non-randomly predicted target events in the range $8.0 \leq M < 8.5$, is conservatively estimated by $\tilde{H} \approx 20\%$.

The M8 algorithm has a modification called M8-MSc, which occasionally reduces the space alarm volume. In that case the requirements to be imposed on the estimates of the measure of space rate of target events become substantially more strict. For this reason we do not examine M8-MSc in this study.

The above analysis of prediction results remains within the framework of a purely scientific (academic) problem, which can be worded as follows: the predictability of large events and its limitations. In this context the question: "Can earthquakes be predicted?" (*Geller et al., 1997*) admits of quantitative answers, e.g., in terms of $\alpha$ and $H$. The answer to the question can become contradictory, when the notion of prediction is made to involve *usefulness* as well. For example, it is rather hard to find a user for the M8 where the spatial alarm unit is 1300 km across. But in case of the M8 non-triviality the size of alarm zone carries important information on the precursory zone size for large earthquake.

Unfortunately, criticisms of prediction often confuse the two criteria. In that case one can advance an argument like the following: if an event has occurred outside the monitored zone, but within the zone of the user's interest, the event is considered unpredicted (see the criticism of CN and M8 for Italy in *Marzocchi, 2008*). Such a broad treatment of prediction results virtually precludes any further statistical analysis of the original method.

The scientific program of the Southern California Earthquake Center focuses on creating *universal* methods for analysis of earthquake forecasting (*Jordan, 2006*; *Zechar et al., 2009*). Moving along this path, we have considered the problem of a fuzzy $\omega(dg)$, which is a tender spot that is common to the analysis of all prediction methods. Whether such difficulties can be overcome may depend on specific features of the algorithm being tested.

# APPENDIX

**Table A.** Sub-regions and centres of the circles for the M8.0+ test.

| N° | Region | Lat | Lon | N° | Region | Lat | Lon | N° | Region | Lat | Lon | N° | Region | Lat | Lon |
|---|---|---|---|---|---|---|---|---|---|---|---|---|---|---|---|
| 1 | TO | -15.00 | -175.00 | 67 | TJ | 31.00 | 132.00 | 133 | CA | 12.00 | -88.00 | 199 | ME | 39.83 | 41.01 |
| 2 | TO | -17.50 | -174.00 | 68 | TJ | 29.00 | 131.00 | 134 | CA | 10.00 | -85.00 | 200 | ME | 39.83 | 51.63 |
| 3 | TO | -20.00 | -175.00 | 69 | TJ | 27.00 | 129.00 | 135 | CA | 8.00 | -82.50 | 201 | ME | 38.93 | 54.81 |
| 4 | TO | -22.50 | -176.00 | 70 | TJ | 26.00 | 127.00 | 136 | CA | 5.00 | -82.50 | 202 | ME | 38.48 | 45.89 |
| 5 | TO | -25.00 | -177.00 | 71 | TJ | 25.00 | 124.50 | 137 | CA9 | 6.00 | -76.00 | 203 | ME | 38.03 | 43.04 |
| 6 | TO | -27.50 | -177.50 | 72 | TJ | 25.00 | 122.00 | 138 | CA9 | 8.00 | -73.00 | 204 | ME | 37.58 | 48.74 |
| 7 | TO | -30.00 | -178.00 | 73 | TJ | 22.50 | 121.50 | 139 | CA9 | 10.00 | -70.00 | 205 | ME | 37.58 | 57.29 |
| 8 | TO | -32.50 | -179.00 | 74 | TJ | 20.00 | 121.50 | 140 | CA9 | 13.00 | -60.00 | 206 | ME | 36.68 | 51.24 |
| 9 | TO | -35.00 | 180.00 | 75 | TJ | 17.50 | 121.00 | 141 | CA9 | 16.00 | -61.00 | 207 | ME | 36.23 | 45.08 |
| 10 | TO | -37.00 | 178.00 | 76 | JB | 45.50 | 150.50 | 142 | CA9 | 19.00 | -63.00 | 208 | ME | 36.23 | 54.04 |
| 11 | TO | -39.00 | 176.00 | 77 | JB | 44.00 | 148.00 | 143 | CA9 | 19.50 | -66.50 | 209 | ME | 35.78 | 59.68 |
| 12 | GU | -2.00 | 136.00 | 78 | JB | 43.50 | 145.50 | 144 | CA9 | 16.00 | -88.00 | 210 | ME | 34.43 | 46.48 |
| 13 | GU | -2.25 | 138.50 | 79 | JB | 43.00 | 143.00 | 145 | SA | 3.00 | -78.00 | 211 | ME | 32.63 | 48.50 |
| 14 | GU | -2.50 | 141.00 | 80 | JB | 41.00 | 141.00 | 146 | SA | 0.00 | -80.00 | 212 | ME | 32.63 | 56.98 |
| 15 | GU | -3.75 | 143.50 | 81 | JB | 39.00 | 142.00 | 147 | SA | -3.00 | -81.50 | 213 | ME | 32.18 | 60.16 |
| 16 | GU | -5.00 | 146.00 | 82 | JB | 36.50 | 141.00 | 148 | SA | -5.00 | -77.00 | 214 | ME | 30.38 | 50.18 |
| 17 | GU | -5.00 | 149.00 | 83 | JB | 35.00 | 139.00 | 149 | SA | -6.00 | -81.00 | 215 | ME | 29.93 | 57.46 |
| 18 | GU | -5.00 | 152.00 | 84 | JB | 33.00 | 141.00 | 150 | SA | -8.00 | -74.00 | 216 | ME | 29.93 | 60.06 |
| 19 | GU | -6.25 | 154.00 | 85 | JB | 31.00 | 142.00 | 151 | SA | -9.00 | -79.00 | 217 | ME | 28.13 | 51.77 |
| 20 | GU | -7.50 | 156.00 | 86 | JB | 29.00 | 142.50 | 152 | SA | -11.00 | -74.00 | 218 | ME | 27.68 | 56.87 |
| 21 | GU | -8.75 | 158.00 | 87 | JB | 27.00 | 143.00 | 153 | SA | -12.00 | -77.50 | 219 | ME | 27.23 | 59.93 |
| 22 | GU | -10.00 | 160.00 | 88 | JB | 25.00 | 143.00 | 154 | SA | -15.00 | -75.00 | 220 | ME | 26.78 | 53.75 |
| 23 | GU | -10.50 | 162.50 | 89 | JB | 23.00 | 143.00 | 155 | SA | -17.50 | -71.00 | 221 | PM | 42.98 | 78.43 |
| 24 | GU | -11.00 | 165.00 | 90 | JB | 21.00 | 144.50 | 156 | SA | -20.50 | -69.00 | 222 | PM | 42.08 | 74.73 |
| 25 | GU | -13.00 | 166.25 | 91 | JB | 19.00 | 146.00 | 157 | SA | -22.00 | -67.00 | 223 | PM | 41.63 | 81.30 |
| 26 | GU | -15.00 | 167.50 | 92 | BM | 16.50 | 147.00 | 158 | SA | -23.50 | -70.00 | 224 | PM | 40.28 | 64.02 |
| 27 | GU | -17.50 | 168.25 | 93 | BM | 14.00 | 146.00 | 159 | SA | -25.00 | -68.00 | 225 | PM | 40.28 | 70.51 |
| 28 | GU | -20.00 | 169.00 | 94 | BM | 12.00 | 144.00 | 160 | SA | -27.00 | -71.00 | 226 | PM | 40.28 | 78.18 |
| 29 | GU | -21.25 | 170.75 | 95 | BM | 12.00 | 141.00 | 161 | SA | -28.00 | -69.00 | 227 | PM | 39.83 | 67.56 |
| 30 | JV | 11.00 | 93.00 | 96 | KK | 46.00 | 152.00 | 162 | SA9 | -30.00 | -71.50 | 228 | PM | 39.83 | 73.46 |
| 31 | JV | 9.50 | 93.75 | 97 | KK | 48.50 | 155.50 | 163 | SA9 | -31.00 | -70.00 | 229 | PM | 39.83 | 75.82 |
| 32 | JV | 7.00 | 94.50 | 98 | KK | 51.00 | 158.00 | 164 | SA9 | -33.00 | -72.50 | 230 | PM | 38.93 | 71.63 |
| 33 | JV | 5.00 | 95.75 | 99 | KK | 53.50 | 160.00 | 165 | SA9 | -34.00 | -71.00 | 231 | PM | 38.48 | 73.82 |
| 34 | JV | 3.00 | 97.00 | 100 | KK | 56.50 | 161.50 | 166 | SA9 | -36.00 | -73.00 | 232 | PM | 38.03 | 69.83 |
| 35 | JV | 2.00 | 98.50 | 101 | AA | 55.00 | 166.50 | 167 | SA9 | -39.00 | -73.50 | 233 | PM | 38.03 | 84.65 |
| 36 | JV | -1.00 | 100.00 | 102 | AA | 53.00 | 171.00 | 168 | SS | -56.00 | -27.00 | 234 | PM | 36.68 | 71.40 |
| 37 | JV | -3.00 | 101.50 | 103 | AA | 51.00 | 176.00 | 169 | SS | -57.00 | -27.00 | 235 | PM | 36.68 | 76.44 |
| 38 | JV | -5.00 | 103.00 | 104 | AA | 51.00 | -178.50 | 170 | SS | -58.50 | -25.50 | 236 | PM | 36.68 | 82.60 |
| 39 | JV | -6.50 | 105.00 | 105 | AA | 51.50 | -173.00 | 171 | MT | 46.13 | 12.68 | 237 | PM | 36.23 | 68.60 |
| 40 | JV | -8.00 | 107.00 | 106 | AA | 52.50 | -167.50 | 172 | MT | 45.68 | 16.32 | 238 | PM | 35.78 | 73.43 |
| 41 | JV | -8.50 | 109.50 | 107 | AA | 54.00 | -162.50 | 173 | MT | 45.23 | 9.92 | 239 | PM | 35.78 | 79.48 |
| 42 | JV | -9.00 | 112.00 | 108 | AA | 55.50 | -157.50 | 174 | MT | 43.43 | 12.09 | 240 | PM | 34.43 | 70.68 |
| 43 | JV | -9.25 | 114.50 | 109 | AA | 56.50 | -152.00 | 175 | MT | 43.43 | 17.05 | 241 | PM | 34.43 | 82.23 |
| 44 | JV | -9.50 | 117.00 | 110 | AA | 60.00 | -153.00 | 176 | MT | 42.08 | 19.22 | 242 | PM | 33.53 | 73.17 |
| 45 | JV | -9.50 | 119.50 | 111 | AA | 63.00 | -151.00 | 177 | MT | 41.63 | 14.10 | 243 | PM | 33.08 | 75.87 |
| 46 | JV | -9.50 | 122.00 | 112 | AA | 62.00 | -145.00 | 178 | MT | 40.73 | 25.67 | 244 | PM | 32.18 | 82.95 |
| 47 | JV | -8.25 | 124.50 | 113 | AA | 60.50 | -140.00 | 179 | MT | 40.73 | 29.21 | 245 | PM | 31.73 | 78.18 |
| 48 | PH | -7.00 | 127.00 | 114 | WA | 47.50 | -122.50 | 180 | MT | 40.73 | 32.75 | 246 | PM | 30.83 | 66.30 |
| 49 | PH | -6.00 | 129.00 | 115 | WA | 44.50 | -130.00 | 181 | MT | 40.28 | 16.23 | 247 | PM | 30.38 | 69.42 |
| 50 | PH | -5.00 | 131.00 | 116 | WA | 43.00 | -126.00 | 182 | MT | 40.28 | 22.72 | 248 | PM | 29.93 | 80.34 |
| 51 | PH | -3.50 | 129.25 | 117 | WA | 40.50 | -128.00 | 183 | MT | 40.28 | 36.29 | 249 | AS | 30.83 | 89.70 |
| 52 | PH | -2.00 | 127.50 | 118 | WA | 40.50 | -123.00 | 184 | MT | 39.83 | 20.36 | 250 | AS | 29.93 | 99.58 |
| 53 | PH | -1.00 | 125.25 | 119 | WA | 38.00 | -119.00 | 185 | MT | 38.93 | 26.97 | 251 | AS | 29.03 | 95.63 |
| 54 | PH | 0.00 | 123.00 | 120 | WA | 37.50 | -122.00 | 186 | MT | 38.48 | 16.25 | 252 | AS | 28.13 | 92.57 |
| 55 | PH | 0.00 | 120.50 | 121 | WA | 35.00 | -118.50 | 187 | MT | 38.48 | 37.91 | 253 | AS | 27.23 | 100.22 |
| 56 | PH | 1.50 | 125.00 | 122 | WA | 37.00 | -118.00 | 188 | MT | 38.03 | 31.07 | 254 | AS | 26.78 | 97.25 |
| 57 | PH | 3.00 | 127.00 | 123 | WA | 32.00 | -115.00 | 189 | MT | 37.58 | 21.38 | 255 | AS | 26.33 | 89.75 |
| 58 | PH | 5.25 | 126.50 | 124 | WA | 29.00 | -112.50 | 190 | MT | 36.68 | 28.28 | 256 | AS | 24.98 | 92.25 |
| 59 | PH | 7.50 | 126.00 | 125 | CA | 23.00 | -108.00 | 191 | MT | 36.23 | 35.56 | 257 | AS | 24.98 | 95.25 |
| 60 | PH | 9.75 | 125.50 | 126 | CA | 20.00 | -109.00 | 192 | MT | 35.33 | 23.38 | 258 | AS | 24.53 | 98.25 |
| 61 | PH | 12.00 | 125.00 | 127 | CA | 18.50 | -106.00 | 193 | MT | 35.33 | 31.08 | 259 | AS | 22.28 | 94.33 |
| 62 | PH | 13.50 | 123.00 | 128 | CA | 19.00 | -103.50 | 194 | MT | 34.88 | 26.68 | 260 | AS | 22.28 | 99.72 |
| 63 | PH | 15.00 | 121.00 | 129 | CA | 17.00 | -100.00 | 195 | MT | 34.43 | 33.28 | 261 | AS | 19.58 | 94.32 |
| 64 | TJ | 35.00 | 136.00 | 130 | CA | 16.00 | -97.00 | 196 | ME | 42.53 | 46.06 | 262 | AS | 16.88 | 94.24 |
| 65 | TJ | 34.00 | 134.00 | 131 | CA | 15.00 | -94.00 | 197 | ME | 41.18 | 48.90 | | | | |
| 66 | TJ | 33.00 | 132.00 | 132 | CA | 14.00 | -91.00 | 198 | ME | 40.73 | 43.96 | | | | |